\documentclass[10pt,aps,preprint]{revtex4-1}

\usepackage{amssymb}
\usepackage{amsmath}
\usepackage{graphics}
\usepackage{latexsym}
\usepackage{array}
\usepackage[activeacute,english]{babel}
\usepackage[latin1]{inputenc}
\usepackage[dvips]{graphicx}
\usepackage{color}
\usepackage{epstopdf}
\usepackage{amssymb}
\usepackage{slashed}
\usepackage{mathrsfs}
\usepackage{amsmath}
\usepackage{wasysym}

\newcommand{\bn}{\begin{eqnarray}}
\newcommand{\en}{\end{eqnarray}}
\newcommand{\be}{\begin{equation}}
\newcommand{\ee}{\end{equation}}

\newcommand{\bc}{\begin{center}}
\newcommand{\ec}{\end{center}}

\begin{document}

\title{\Large \bf Quantum phase transition in the chirality of the (2+1)-dimensional Dirac oscillator}

\author{C. Quimbay\footnote{Associate researcher of Centro
Internacional de F\'{\i}sica, Bogot\'{a} D.C., Colombia.}}
\email{cjquimbayh@unal.edu.co}

\affiliation{Departamento de F\'{\i}sica, Universidad Nacional de Colombia.\\
Ciudad Universitaria, Bogot\'{a} D.C., Colombia.}

\author{P. Strange}
\email{P.Strange@kent.ac.uk}

\affiliation{School of Physical Science, University of Kent.\\
Canterbury, United Kingdom.}

\date{\today}

\begin{abstract}
We study the (2+1)-dimensional Dirac oscillator in the presence of an
external uniform magnetic field ($B$). We show how the change of
the strength of $B$ leads to the existence of a quantum phase
transition in the chirality of the system. A critical value of the
strength of the external magnetic field ($B_c$) can be naturally
defined in terms of physical parameters of the system.
While for $B=B_c$ the fermion can be considered as a free particle
without defined chirality, for $B<B_c$ ($B>B_c$) the chirality is
left (right) and there exist a net potential acting on the
fermion. For the three regimes defined in the quantum phase
transition of chirality, we observe that the energy spectra for
each regime is drastically different. Then, we consider the
$z$-component of the orbital angular momentum as an order
parameter that characterizes the quantum phase transition. \\

\vspace{0.4cm} \noindent {\it{Keywords:}} {Dirac oscillator in
(2+1) dimensions, quantum phase transition, chirality, orbital
angular momentum.}

\vspace{0.4cm} \noindent {\it{PACS Codes:}}  03.65.Pm, 72.80.Vp,
71.70.Di.

\end{abstract}

\maketitle

\section{Introduction}
The system constituted by a relativistic fermion under the action
of a linear vector potential is called the Dirac oscillator
\cite{MS89}. One the most interesting characteristic of the Dirac
oscillator equation is that when its non-relativistic limit is
taken, the associated Klein-Gordon equations describe an harmonic
oscillator in the presence of a strong spin-orbit coupling
\cite{ST98}. Another characteristic of the Dirac oscillator
equation is that it can be exactly solved in one, two and three
dimensions. This is the reason why this system has been one of the
most popular in the context of the relativistic quantum mechanics
and mathematical physics during the last twenty years
\cite{MM89}-\cite{BA2013}. Fortunately, as stimulation to the
search for physical applications of the Dirac oscillator, a first
experimental realization of this system was reported recently
\cite{FV2013}.

On the other hand, some of the properties of the (2+1)-dimensional
Dirac oscillator equation have been studied with the purpose of
establishing a connection between this system and the
(Anti)-Jaynes-Cummings model of the quantum optics
\cite{AB07-1}-\cite{BO2013}. In particular, by defining the chiral
annihilation and creation operators, it has been possible to
calculate the energy spectra of the Dirac oscillator without and
with the presence of a uniform external magnetic field
\cite{AB07-1}-\cite{AB082}. Using the same technique it has been
shown that the linear term in the Dirac oscillator leads directly
to a description of a fermion with an intrinsic left chirality,
which is only present in the two-dimensional Dirac oscillator
\cite{AB07-1}. Additionally, with the same technique, it has been
shown that the system constituted by the two-dimensional Dirac
oscillator in presence of an uniform external magnetic field
presents a chiral quantum phase transition \cite{AB082}. Some of
the characteristics of this quantum phase transition are known
\cite{AB082}, however a study of how the energy spectra and
orbital angular momentum change exactly in each regimes defined in
this phase transition is lacking, nor has it been studied how
these two physical observable change when the non-relativistic
limit is taken.

The main goal of this work is to show precisely how the energy
spectra and the $z$-component of orbital angular momentum change
during the quantum phase transition in the chirality, which is
defined for the (2+1)-dimensional Dirac oscillator in presence of
an external uniform magnetic field ($B$). We show how these
changes both for the relativistic case and in the non-relativistic
limit. We can show precisely how these physical observables change
during the quantum phase transition because we introduce a new
exact technique that employs the number operators of right and
left chirality. In this work, the change in the strength of $B$
permits us to see how the quantum phase transition in the
chirality is present in the system. To do this, we first identify
a critical value of strength of the external magnetic field
($B_c$). The value of $B_c$ can be naturally defined in terms of
physical parameters involved in the system, because for the regime
$B=B_c$ the fermion can be considered as a free particle without
defined chirality. Then using the technique of the number
operators of right and left chirality, we show that for the regime
$B<B_c$ the fermion has left chirality, while for the regime
$B>B_c$ the fermion has right chirality. For the three regimes
associated to the quantum phase transition, we observe that the
energy spectra in each regime is drastically different. Using the
energy spectra previously obtained, we consider the $z$-component
of the orbital angular momentum as an order parameter that
characterizes the quantum phase transition.

The structure of this paper is the following. First, in the
section II, for the Dirac oscillator in presence of an external
uniform magnetic field, we obtain two coupled equations that
permit us to define the regime of critical external magnetic field.
Next, in the section III, by introducing the right and left
chirality number operators, we study the regime of weak
external magnetic fields which define the left chirality phase.
Then, in section IV, we study the regime of strong external
magnetic fields which define the right chirality phase.
In section V, we show that the energy spectra are a signature
of the quantum phase transition in the chirality in both the
relativistic case and the non-relativistic limit. Next, in
section VI, we consider the z-component of the orbital angular
momentum as an order parameter, showing its value for each of
the two phases and how at the critical point it is an undefined
parameter. Finally, in section VII we present some conclusions.

\section{Regime of critical external magnetic field}

The system constituted by a (2+1)-dimensional Dirac oscillator in
presence of an external uniform magnetic field is described by the
following equation
\begin{align}\label{eq:dem1}
i\hbar \frac{\partial}{\partial t} |{\bf \psi}\rangle =\left[c
\sum_{i=1}^{2}\sigma_j\left(p_j - im\omega\sigma_z r_j +eA_j
\right) + \sigma_z m c^2 \right]|{\bf \psi}\rangle,
\end{align}
where $c$ is the light speed, $m$ is the fermion mass, $\omega$ is
the Dirac oscillator frequency, $A_j$ are the components of the
vector potential, $p_j$ are the components of the linear
momentum, $r_j$ are the spatial coordinates in the $(x,y)$ plane
with respect to the origin of the potential, $\sigma_j$ are the
non-diagonal Pauli matrices, $\sigma_z$ is the diagonal Pauli
matrix. The spinor $|\psi\rangle$ is written as
\begin{align}\label{eq:esDir}
|\psi\rangle =\begin{pmatrix}
              |\psi_1\rangle\\
              |\psi_2\rangle
            \end{pmatrix}\,\exp(-iEt/\hbar),
\end{align}

The external uniform magnetic field acting over the
(2+1)-dimensional Dirac oscillator is defined perpendicular to the
plane in the form $\vec B = - B \hat e_z$. The associated vector
potential $\vec A$, then has the form $\vec
A=(A_x,A_y,A_z)=\frac{B}{2}(y,-x,0)$. Substituting the spinor
expressed by (\ref{eq:esDir}) in Eq. (\ref{eq:dem1}) and using the
explicit form of the Pauli matrices $\sigma_x$, $\sigma_y$ and
$\sigma_z$, we obtain the following two coupled equations
\begin{align}
(E-mc^2) |\psi_1\rangle &= c \left[(p_x + im\omega x
-im\tilde{\omega} x)- i(p_y + im\omega y -im\tilde{\omega}
y)\right] |\psi_2\rangle, \label{eq:de11}\\
(E+mc^2) |\psi_2\rangle &= c \left[(p_x - im\omega x
+im\tilde{\omega} x)+ i(p_y - im\omega y +im\tilde{\omega}
y)\right] |\psi_1\rangle,\label{eq:de12}
\end{align}
where $\tilde{\omega}=\omega_c/2$, with $\omega_c=eB/m$ denoting
the cyclotron frequency. For the case in which
$\omega=\tilde{\omega}$, {\it i.e.} $\omega_c=2\omega$, the Eqs.
(\ref{eq:de11}) and (\ref{eq:de12}) lead to
\begin{align}
(E-mc^2) |\psi_1\rangle &= c \left[p_x - i p_y \right]
|\psi_2\rangle, \label{eq:de13}\\
(E+mc^2) |\psi_2\rangle &= c \left[p_x + i p_y \right]
|\psi_1\rangle,\label{eq:de14}
\end{align}
and Eq. (\ref{eq:dem1}) then becomes
\begin{align}\label{eq:dem2}
i\hbar \frac{\partial}{\partial t} |{\bf \psi}\rangle =\left[c
\sum_{i=1}^{2}\sigma_j p_j + \sigma_z m c^2 \right]|{\bf
\psi}\rangle,
\end{align}
that corresponds to the (2+1)-dimensional Dirac equation
describing a relativistic fermion of mass $m$ which moves freely
over the $xy$-plane. Starting from Eqs. (\ref{eq:de13}) and
(\ref{eq:de14}), we substitute one into the other and we obtain
the associated Klein-Gordon equations
\begin{align}
(E^2-m^2c^4) \mid \psi_1 \rangle &= c^2 p^2 \mid \psi_1\rangle, \label{eq:de13a}\\
(E^2-m^2c^4) \mid \psi_2 \rangle &= c^2 p^2 \mid \psi_2\rangle,
\label{eq:de14a}
\end{align}
where $p^2 = p_x^2 + p_y^2$. Therefore, the Eqs. (\ref{eq:de13a})
and (\ref{eq:de14a}) are describing a same free fermion with
relativistic energy given by
\begin{align}
E= \pm \sqrt{\hbar^2k^2c^2+ m^2c^4}, \label{eq:enes0}
\end{align}
where we have written $p^2=\hbar^2k^2$, with $k$ being the wavevector
associated with the momentum $p$. This means that for the case in
which the strength of an external magnetic field satisfies the
critical condition $\omega_c=eB_c/m=2\omega$, or equivalently
\begin{align}\label{eq:cmf}
B_c=2m\omega/e,
\end{align}
then the effect of the linear potential $im\omega\sigma_z r_j$ on
the fermion in Eq. (\ref{eq:dem1}) is annulled by the effect of
the vector potential $eA_j$. In this form, it is possible to
speculate that the quantity $m\omega$ that appears in the linear
potential of the Dirac oscillator defined by Eq. (\ref{eq:de1})
can be written as $m\omega=e B_I/2$, where $B_I$ can be
interpreted as an effective magnetic field acting on the fermion.
In other words, the effect of the linear potential on the fermion
described by the Dirac oscillator is equivalent to the effect that
an uniform magnetic field $\vec B_I = B_I \hat e_z$ has on the
fermion. Therefore, for the case $\omega=\tilde{\omega}$, the
effective magnetic field $\vec B_I = B_I \hat e_z$ is canceled by
the external magnetic field $\vec B_c = -B_I \hat e_z$, and the
Eq. (\ref{eq:dem1}) describes a free fermion.

If the external magnetic field vanishes in Eq. (\ref{eq:dem1}),
this means that $A_j=0$, then we obtain the (2+1)-dimensional
Dirac oscillator equation given by
\begin{align}\label{eq:de1}
i\hbar \frac{\partial}{\partial t} |{\bf \psi}\rangle =\left[c
\sum_{i=1}^{2}\sigma_j\left(p_j - im\omega\sigma_z r_j \right) +
\sigma_z m c^2 \right]|{\bf \psi}\rangle.
\end{align}
This equation has been extensively studied and many of its
properties are very well known \cite{AB07-1}-\cite{BO2013},
\cite{QUI2013}. After the Eq. (\ref{eq:de1}) is solved, the energy
spectrum is given by \cite{AB07-1}
\begin{align}
E_{n_l}= \pm \sqrt{\hbar^2k_{n_l}^2c^2 + m^2c^4}, \label{eq:enes1}
\end{align}
where the wave length $k_{n_l}$ has been defined as
\begin{align}
k_{n_l}^2=k^2\left(n_l + \frac{1}{2}\mp \frac{1}{2}\right),
\label{eq:qlw}
\end{align}
with $k^2=4m\omega/\hbar$ and $n_l=0,1,2,\ldots$. We can observe
the energy spectrum (\ref{eq:enes1}) has the same form that the
energy for the case of free fermion (\ref{eq:enes0}), but now the
wave length takes discrete values as is shown in (\ref{eq:qlw}).
Thus, we can observe that the main effect of the linear potential
that appears in Eq. (\ref{eq:de1}) is to quantize the energy
spectrum.

The strength of an external magnetic field $B$, in relation with
the value of the critical magnetic field $B_c$, is the parameter
that will determine the chirality of the states describe by the
Eq. (\ref{eq:dem1}). In the next two sections, we will solve the
coupled system given by Eqs. (\ref{eq:de11}) and (\ref{eq:de12})
for two different regimes: (i) weak external magnetic fields
$\omega > {\tilde \omega}$, or $B<B_c$; (ii) strong external
magnetic fields $\omega < {\tilde \omega}$, or $B>B_c$.

\section{Regime $\omega - {\tilde \omega} > 0$: Weak external magnetic fields}

If the external magnetic field satisfies the condition $B<B_c$,
{\it i.e.} $\omega - {\tilde \omega} > 0$, then Eqs.
(\ref{eq:de11}) and (\ref{eq:de12}) can be written as
\begin{align}
(E-mc^2) |\psi_1\rangle &= c \left[(p_x + im\omega_T x)- i(p_y +
im\omega_T y)\right] |\psi_2\rangle, \label{eq:de15}\\
(E+mc^2) |\psi_2\rangle &= c \left[(p_x - im\omega_T x)+ i(p_y -
im\omega_T y)\right] |\psi_1\rangle,\label{eq:de16}
\end{align}
where
\begin{align}
\omega_T=\omega - {\tilde \omega}.
\end{align}
Using Eqs. (\ref{eq:de15}) and (\ref{eq:de16}), it is possible to
substitute one into the other and obtain the associated
Klein-Gordon equations
\begin{align}
(E^2-m^2c^4) |\psi_1\rangle &= 2mc^2 \left[H_{ho}
-\hbar\omega_T-\omega_T L_z\right] |\psi_1\rangle, \label{eq:de17}\\
(E^2-m^2c^4) |\psi_2\rangle &= 2mc^2 \left[H_{ho}
+\hbar\omega_T-\omega_T L_z\right] |\psi_2\rangle,\label{eq:de18}
\end{align}
where we have used the quantum mechanics commutation relations
$[x,p_x]=[y,p_y]=i\hbar$, $[x,y]=[p_x,p_y]=0$. In Eqs.
(\ref{eq:de17}) and (\ref{eq:de18}), $H_{ho}$ and $L_z$ represent
respectively the two-dimensional harmonic oscillator Hamiltonian
and the $z$-component of the orbital angular momentum
\begin{align}
H_{ho}&= \frac{{\bf p}^2}{2m} +\frac{m\omega_T^2}{2}{\bf
r}^2,\label{eq:hoh}\\
L_z&=x p_y-p_y x,\label{eq:amo}
\end{align}
where ${\bf r}^2=x^2+y^2$ and ${\bf p}^2=p_x^2+p_y^2$. Now we
introduce the right chiral annihilation and creation operators
given respectively by $a_r =\frac{1}{\sqrt 2}(a_x-ia_y)$,
$a_r^\dag =\frac{1}{\sqrt 2}(a_x^\dag+ia_y^\dag)$ and the left
chiral annihilation and creation operators given respectively by
$a_l =\frac{1}{\sqrt 2}(a_x+ia_y)$, $a_l^\dag =\frac{1}{\sqrt
2}(a_x^\dag-ia_y^\dag)$, where $a_x$, $a_y$, $a_x^\dag$ and
$a_y^\dag$ are the usual annihilation and creation operators of
the harmonic oscillator defined respectively as $a_j
=\frac{1}{\sqrt 2}\left(\frac{1}{\Delta} r_j +
i\frac{\Delta}{\hbar} p_j\right)$ and $a_j^\dag =\frac{1}{\sqrt
2}\left(\frac{1}{\Delta} r_j - i\frac{\Delta}{\hbar} p_j\right)$,
with $\Delta = v_f/\omega$ representing the ground-state
oscillator width. Because the orbital angular momentum $L_z$ is
written in terms of the right and left chiral operators as
$L_z=\hbar(a_r^\dag a_r - a_l^\dag a_l)$, then $a_r^\dag$
($a_l^\dag$) is interpreted as the operator that create a right
(left) quantum of angular momentum \cite{AB07-1}. If we introduce
the number operators of right and left chirality $N_r$ and $N_l$
defined by
\begin{align}
N_r=a_r^\dag a_r, \label{eq:lno}\\
N_l=a_l^\dag a_l, \label{eq:rno}
\end{align}
then expressions (\ref{eq:hoh}) and (\ref{eq:amo}) become
\cite{Cohen1991}
\begin{align}
H_{ho}&= \hbar \omega_T (N_r + N_l +1),\label{eq:hoh1}\\
L_z&=\hbar (N_r - N_l).\label{eq:amo1}
\end{align}
If we substitute expressions (\ref{eq:hoh1}) and (\ref{eq:amo1})
in Eqs. (\ref{eq:de17}) and (\ref{eq:de18}), we obtain
\begin{align}
(E^2-m^2c^4) |\psi_1\rangle &= 4mc^2\hbar\omega_T N_l
|\psi_1\rangle, \label{eq:de19}\\
(E^2-m^2c^4) |\psi_2\rangle &= 4mc^2\hbar\omega_T(N_l+1)
|\psi_2\rangle,\label{eq:de20}
\end{align}
where we observe that only the left chiral number operator $N_l$
appears, implying that for the regime $B<B_c$ the eigenstates
$|+E_{n_l}\rangle$ and $|-E_{n_l}\rangle$ describe quantum
states of left chirality. From Eqs. (\ref{eq:de19}) and
(\ref{eq:de20}) and working in the base of eigenstates of the
number operator $N_l$, we obtain the energy spectrum for fermions
of left chirality
\begin{align}
E_{n_l}= \pm \sqrt{F_w\hbar^2k_{n_l}^2c^2 + m^2c^4},
\label{eq:enes2}
\end{align}
where the function $F_w$ is given by
\begin{align}
F_w=1-\frac{eB}{2m\omega},
\end{align}
and $k_{n_l}^2=k^2\left(n_l + \frac{1}{2}\mp \frac{1}{2}\right)$,
with $k^2=4m\omega/\hbar$ and $n_l=0,1,2,\ldots$.

For the case in which the external magnetic field vanishes
$(B=0)$, then $F_w=1$ and the energy spectrum (\ref{eq:enes2})
leads consistently to the energy spectrum of the Dirac oscillator
given by the expression (\ref{eq:enes1}).

Now we will obtain the energy spectrum in the non-relativistic
limit. For the case $E>0$, the non-relativistic limit is taken by
means of the approximation
\begin{align}
E^2-m^2c^4 \simeq 2mc^2 E^{+}. \label{eq:nrlpe}
\end{align}
Substituting (\ref{eq:nrlpe}) in (\ref{eq:de19}), we obtain that
the energy spectrum is
\begin{align}
E^{+}_{n_l}= F_w\frac{\hbar^2k^2}{2m} n_l, \label{eq:enes3}
\end{align}
with $n_l=0,1,2,\ldots$. For the case $E<0$, the non-relativistic
limit is taken by means of the approximation
\begin{align}
E^2-m^2c^4 \simeq -2mc^2 E^{-}. \label{eq:nrlne}
\end{align}
Substituting (\ref{eq:nrlne}) in (\ref{eq:de20}), we obtain that
the energy spectrum is
\begin{align}
E^{-}_{n_l}= -F_w\frac{\hbar^2k^2}{2m}(n_l+1), \label{eq:enes4}
\end{align}
with $n_l=0,1,2,\ldots$.

\section{Regime ${\tilde \omega} - \omega > 0$: Strong external magnetic fields}

If the external magnetic field satisfies the condition $B>B_c$,
{\it i.e.} ${\tilde \omega} - \omega > 0$, then the Eqs.
(\ref{eq:de11}) and (\ref{eq:de12}) can be written as
\begin{align}
(E-mc^2) |\psi_1\rangle &= c \left[(p_x - im{\tilde \omega}_T x)-
i(p_y -
im{\tilde \omega}_T y)\right] |\psi_2\rangle, \label{eq:de21}\\
(E+mc^2) |\psi_2\rangle &= c \left[(p_x + im{\tilde \omega}_T x)+
i(p_y + im{\tilde \omega}_T y)\right]
|\psi_1\rangle,\label{eq:de22}
\end{align}
where
\begin{align}
{\tilde \omega}_T={\tilde \omega}-\omega.
\end{align}
From Eqs. (\ref{eq:de21}) and (\ref{eq:de22}), we substitute one
into the other and we obtain the associated Klein-Gordon equations
\begin{align}
(E^2-m^2c^4) |\psi_1\rangle &= 2mc^2 \left[{\tilde H}_{ho}
+\hbar{\tilde \omega}_T+{\tilde \omega} L_z\right]
|\psi_1\rangle, \label{eq:de23}\\
(E^2-m^2c^4) |\psi_2\rangle &= 2mc^2 \left[{\tilde H}_{ho}
-\hbar{\tilde \omega}_T+{\tilde \omega} L_z\right]
|\psi_2\rangle,\label{eq:de24}
\end{align}
where we have used the usual quantum mechanics commutation
relations and $L_z$ is given by the expression (\ref{eq:amo}). In
Eqs. (\ref{eq:de23}) and (\ref{eq:de24}), ${\tilde H}_{ho}$ is the
two-dimensional harmonic oscillator Hamiltonian described by
\begin{align}
{\tilde H}_{ho}&= \frac{{\bf p}^2}{2m} +\frac{m{\tilde
\omega}_T^2}{2}{\bf r}^2. \label{eq:hoh2}
\end{align}
We again introduce the number operators of right and left
chirality $N_r$ and $N_l$, respectively. In terms of these
operators, $L_z$ is written as expression (\ref{eq:amo1}),
while ${\tilde H}_{ho}$ is
\begin{align}
{\tilde H}_{ho}= \hbar {\tilde \omega}_T (N_r + N_l
+1),\label{eq:hoh3}
\end{align}
After expressions (\ref{eq:hoh3}) and (\ref{eq:amo1}) are
substituted in Eqs. (\ref{eq:de23}) and (\ref{eq:de24}), we obtain
\begin{align}
(E^2-m^2c^4) |\psi_1\rangle &= 4mc^2\hbar{\tilde \omega}_T (N_r+1)
|\psi_1\rangle, \label{eq:de25}\\
(E^2-m^2c^4) |\psi_2\rangle &= 4mc^2\hbar{\tilde \omega}_T N_r
|\psi_2\rangle,\label{eq:de26}
\end{align}
where we now observe that only the right chiral number
operator $N_r$ appears, implying that for the regime $B>B_c$ the
eigenstates $|+E_{n_r}\rangle$ and $|-E_{n_r}\rangle$ describe
quantum states of right chirality. From Eqs.
(\ref{eq:de25}) and (\ref{eq:de26}) and working in the base of
eigenstates of the number operator $N_r$, we obtain the energy
spectrum for fermions of right chirality
\begin{align}
E_{n_r}= \pm \sqrt{F_s2\hbar c^2 e B \left( n_r + \frac{1}{2} \mp
\frac{1}{2}\right)+ m^2c^4}, \label{eq:enes5}
\end{align}
where the function $F_s$ is given by
\begin{align}
F_s=1-\frac{2m\omega}{eB},
\end{align}
and $n_r=0,1,2,\ldots$.

For the case in which the linear potential in Eq. (\ref{eq:dem1})
vanishes $(\omega=0)$, the system that we are considering
corresponds to a massive fermion having electric charge in
presence of a uniform magnetic field $B$. For this case, $F_s=1$
and the energy spectrum (\ref{eq:enes5}) is written as
\begin{align}
E_{n_r}= \pm \sqrt{2\hbar c^2 e B \left( n_r + \frac{1}{2} \pm
\frac{1}{2}\right)+ m^2c^4}, \label{eq:enes6}
\end{align}
with $n_r=0,1,2,\ldots$, which corresponds to the Landau-level
spectrum.

For the case $E>0$, we find that the energy spectrum in the
non-relativistic limit is
\begin{align}
E^{+}_{n_r}= F_s\frac{\hbar eB}{m} (n_r +1), \label{eq:enes7}
\end{align}
while for the case $E<0$, we find that this spectrum is
\begin{align}
E^{-}_{n_r}= -F_s\frac{\hbar eB}{m} n_r. \label{eq:enes8}
\end{align}

\section{Energy spectra as signature of the quantum phase transition}

We have shown in the previous sections that the chirality of the
fermion described by Eq. (\ref{eq:dem1}) changes drastically
accordingly to the strength of the external magnetic field $B$
relative to the critical value $B_c$. To show this, we have
defined three different regimes: (i) weak external magnetic fields
($B<B_c$); (ii) a critical external magnetic field ($B=B_c$);
(iii) strong external magnetic fields ($B>B_c$). The change of
chirality that we have observed is: For $B<B_c$, the fermion has
left chirality; for $B=B_c$, the fermion has not a defined
chirality; for $B>B_c$, the fermion has right chirality.

In this section we will show that the energy is a good signature
with which to characterize the quantum phase transition in the
chirality, because the energy spectrum of the fermion changes
drastically with the regime considered. We will show also that
that the origin of this phase transition of chirality is not a
consequence of relativistic effects because the changes in the
energy spectrum appear also in the non-relativistic limit.

For $B<B_c$, from the energy spectrum (\ref{eq:enes2}), we obtain
that the relativistic discrete energy spectra are
\begin{align}
E_{n_l}= \pm \sqrt{4\hbar c^2 m\omega \left(1-\frac{eB}{2m\omega}
\right)\left(n_l + \frac{1}{2}\mp
\frac{1}{2}\right)},\label{eq:enes10}
\end{align}
with $n_l=0,1,2,\ldots$. For $B=B_c$, from (\ref{eq:enes0}), the
relativistic continuum energy spectra are
\begin{align}
E= \pm \sqrt{p^2c^2+ m^2c^4}. \label{eq:enes11}
\end{align}
For $B>B_c$, from the energy spectrum (\ref{eq:enes5}), the
relativistic discrete energy spectra are
\begin{align}
E_{n_r}= \pm \sqrt{2\hbar c^2 e B \left( 1-\frac{2m\omega}{eB}
\right) \left( n_r + \frac{1}{2} \pm \frac{1}{2}\right)+ m^2c^4},
\label{eq:enes12}
\end{align}
with $n_r=0,1,2,\ldots$. If $B \ll B_c$, {\it i.e.} $B \simeq 0$,
from (\ref{eq:enes11}) we obtain the energy spectra of the
(2+1)-dimensional Dirac oscillator \cite{QUI2013}
\begin{align}
E_{n_l}= \pm \sqrt{4\hbar c^2 m\omega \left(n_l + \frac{1}{2}\mp
\frac{1}{2}\right)}.\label{eq:enes13}
\end{align}
with $n_l=0,1,2,\ldots$. If $B \gg B_c$, {\it i.e.} $\omega \simeq
0$, from (\ref{eq:enes13}) we obtain the Landau-level spectra
\cite{QUI2013}
\begin{align}
E_{n_r}= \pm \sqrt{2\hbar c^2 e B \left( n_r + \frac{1}{2} \mp
\frac{1}{2}\right)+ m^2c^4}, \label{eq:enes14}
\end{align}
with $n_r=0,1,2,\ldots$. We observe that the relativistic energy
spectra for the three regimes are very different.

For $B<B_c$, we rewrite the expressions (\ref{eq:enes3}) and
(\ref{eq:enes4}) and we obtain the non-relativistic discrete
spectra for positive and negative energies
\begin{align}
E^{+}_{n_l}&= \frac{\hbar^2k^2}{2m} \left(1-\frac{eB}{2m\omega}
\right) n_l, \label{eq:enes15}\\
E^{-}_{n_l}&= -\frac{\hbar^2k^2}{2m}\left(1-\frac{eB}{2m\omega}
\right)(n_l+1), \label{eq:enes16}
\end{align}
respectively, with $n_l=0,1,2,\ldots$. For $B=B_c$, the
non-relativistic continuum spectra for positive and negative
energies are
\begin{align}
E^{\pm}= \pm \frac{p^2}{2m^2}. \label{eq:enes17}
\end{align}
For $B>B_c$, we rewrite the expressions (\ref{eq:enes7}) and
(\ref{eq:enes8}) and we obtain the non-relativistic discrete
spectra for positive and negative energies
\begin{align}
E^{+}_{n_r}&= 2\hbar c^2 e B \left(1-\frac{2m\omega}{eB}
\right) (n_r +1), \label{eq:enes18}\\
E^{-}_{n_r}&= -2\hbar c^2 e B \left(1-\frac{2m\omega}{eB}
\right)n_r, \label{eq:enes19}
\end{align}
respectively, with $n_r=0,1,2,\ldots$.

As happens in the relativistic case, the non-relativistic limit
the energy spectra are very different in the three regimes. This
means that the phase transition in the chirality does not have a
relativistic origin. This quantum phase transition exists in the
relativistic and non-relativistic cases.

\section{Orbital angular momentum as an order parameter}

Now we will consider the $z$-component of the orbital angular
momentum $L_z$ as an order parameter that characterizes the
quantum phase transition in the chirality. We will show that the
physical magnitude of this quantity takes on different values in
the two phases and is undetermined in the critical point. For
$B<B_c$, the system is in the left chirality phase and the
expectation value of $L_z$ denoted as $<L_z>$ is negative,
{\it i.e.} $<L_z> <0$ , while for $B>B_c$, the system is in the
right chirality phase and $<L_z> >0$.

For the left chirality phase, {\it i.e.} for $B<B_c$, from the
associated Klein-Gordon equations given by Eqs. (\ref{eq:de17})
and (\ref{eq:de18}), and using the expressions (\ref{eq:hoh1}) and
(\ref{eq:amo1}), we obtain
\begin{align}
\langle L_z^+\rangle_{n_l} = \langle +E_{n_l}|L_z|+E_{n_l}\rangle
&=-\frac{E_{n_l}^2-m^2c^4}{2c^2m\omega_T}+ \hbar n_l, \label{eq:de31}\\
\langle L_z^-\rangle_{n_l} = \langle -E_{n_l}|L_z|-E_{n_l}\rangle
&= -\frac{E_{n_l}^2-m^2c^4}{2c^2m\omega_T}+ \hbar
(n_l+2),\label{eq:de32}
\end{align}
with $n_l=0,1,2,\ldots$. For the right chirality phase, {\it i.e.}
for $B>B_c$, from the associated Klein-Gordon equations given by
Eqs. (\ref{eq:de23}) and (\ref{eq:de24}), and using the
expressions (\ref{eq:hoh3}) and (\ref{eq:amo1}), we obtain
\begin{align}
\langle L_z^+\rangle_{n_r} = \langle +E_{n_r}|L_z|+E_{n_r}\rangle
&=\frac{E_{n_r}^2-m^2c^4}{2c^2m{\tilde \omega}_T}- \hbar
(n_r +2), \label{eq:de33}\\
\langle L_z^-\rangle_{n_r} = \langle -E_{n_r}|L_z|-E_{n_r}\rangle
&= \frac{E_{n_r}^2-m^2c^4}{2c^2m{\tilde \omega}_T}- \hbar
n_r,\label{eq:de34}
\end{align}
with $n_r=0,1,2,\ldots$. For the critical point, {\it i.e.} for
$B=B_c$, from the associated Klein-Gordon equations given by Eqs.
(\ref{eq:de13a}) and (\ref{eq:de14a}), we observe that $\langle
L_z\rangle$ is undetermined.

The corresponding expectation values of the orbital angular
momentum in the non-relativistic limit are the following. For the
left chirality phase ($B<B_c$)
\begin{align}
\langle L_z^+\rangle_{n_l} = \langle +E_{n_l}|L_z|+E_{n_l}\rangle
&=-\frac{1}{\omega_T}E_{n_l}+ \hbar n_l, \label{eq:de35}\\
\langle L_z^-\rangle_{n_l} = \langle -E_{n_l}|L_z|-E_{n_l}\rangle
&= \frac{1}{\omega_T}E_{n_l}+ \hbar (n_l+2),\label{eq:de36}
\end{align}
with $n_l=0,1,2,\ldots$. For the right chirality phase ($B>B_c$)
\begin{align}
\langle L_z^+\rangle_{n_r} = \langle +E_{n_r}|L_z|+E_{n_r}\rangle
&=\frac{1}{{\tilde \omega}_T}E_{n_r}- \hbar (n_r+2), \label{eq:de35}\\
\langle L_z^-\rangle_{n_r} = \langle -E_{n_r}|L_z|-E_{n_r}\rangle
&= -\frac{1}{{\tilde \omega}_T}E_{n_r}-\hbar n_r,\label{eq:de36}
\end{align}
with $n_r=0,1,2,\ldots$. For the critical point $B=B_c$, these
values are undetermined.

As happens in the relativistic case, in the non-relativistic limit
the expectation values of the orbital angular momentum are drastically
different between the two phases and are undetermined in the critical
point.

\section{Conclusions}

We have shown how the energy spectra and the $z$-component of the
orbital angular momentum change exactly during the quantum phase
transition of chirality which is defined for the (2+1)-dimensional
Dirac oscillator in presence of an external uniform magnetic field
($B$). These changes have been obtained both in the relativistic
case and in the non-relativistic limit. After introducing a new
technique that works with the number operators of right and left
chirality, we have performed an explicit and exact calculation to
show how these physical observables change during the quantum phase
transition. We have demonstrated that the changing the strength of
$B$ leads to a quantum phase transition in the chirality of this system.

We have first identified a critical value of strength of the
external magnetic field ($B_c$) which is naturally defined in
terms of physical parameters involved in the system, because for
the regime $B=B_c$ the fermion can be considered as a free
particle without defined chirality. We have used the number
operator technique to show that for the regime $B<B_c$ the fermion
has left chirality, while for the regime $B>B_c$ the fermion has
right chirality. We have observed that, for the three regimes
associated with the quantum phase transition, the energy spectra
in each of the three regimes is drastically different. By using
the energy spectra previously obtained, we have considered the
$z$-component of the orbital angular momentum as an order parameter
that characterizes the quantum phase transition.

The advance made in this work has been to characterize the quantum
phase transition in the chirality in an exact form. The new
technique, which uses the number operators for right and left
chirality, that we have introduced in this work has permitted us
to calculate exactly the energy spectra and the z-component of the
angular momentum. A possible application of the results that we
have presented in this work can be in the context of the
description of the electronic properties of monolayer and bilayer
graphene \cite{QUI2013}.

\section*{Acknowledgments}

C. Quimbay thanks the School of Physical Sciences at the
University of Kent, Canterbury, for their hospitality during his
visit there.


\end{document}